\documentstyle[preprint,aps,prd]{revtex}
\begin{document}
\title{Quantum inequalities for the free \\Rarita-Schwinger fields in flat spacetime}
\author{ Hongwei Yu $^{a,b}$ and Puxun Wu $^b$ }

\address
{ $^a $CCAST(World Lab.), P. O. Box 8730, Beijing, 100080, P. R.
China.\\
$ ^b$Department of Physics and Institute of  Physics,\\ Hunan
Normal University, Changsha, Hunan 410081, China\footnote{Mailing
address} .}
\tightenlines \maketitle
\begin{abstract}
Using the methods developed by Fewster and colleagues, we derive a
quantum inequality for the free massive spin-${3\over 2}$
Rarita-Schwinger fields in the four dimensional Minkowski
spacetime. Our quantum inequality bound for the Rarita-Schwinger
fields is weaker, by a factor of 2, than that for the
spin-${1\over 2}$ Dirac fields. This fact along with other quantum
inequalities obtained by various other authors for the fields of
integer spin (bosonic fields) using similar methods lead us to
conjecture that, in the flat spacetime,  separately for bosonic
and fermionic fields, the quantum inequality bound gets weaker as
the the number of degrees of freedom of the field increases. A
plausible physical reason might be that the more the number of
field degrees of freedom, the more freedom one has to create
negative energy, therefore, the weaker the quantum inequality
bound.

\vspace{0.2cm}
\end{abstract}

\newpage

\section{Introduction}

It is well established that the energy density of a field, which
is strictly positive in classical physics, can become negative and
even unbounded from below in quantum field theory due to quantum
coherence effects\cite{EGJ}. Specific experimentally studied
examples of quantum states exhibiting negative energy density are
squeezed states of light in quantum optics\cite{WKHW} and the
Casimir vacuum state of quantized fields\cite{LMR}. As a result,
all the known pointwise energy conditions in classical general
relativity, such as the weak energy condition and null energy
condition, can be violated.

However, if the negative energy density in quantum field theory is
unconstrained, i.e., if an arbitrary amount of negative energy is
allowed to persist for an arbitrary long period of time, then
serious ramifications result. These include exotic phenomena such
as violation of the second law of thermodynamics\cite{L1,7},
traversable wormholes\cite{10,11}, "warp drive"\cite{12}, and even
time machines\cite{11,13}. Therefore, a lot of effort has been
made toward determining the extent to which these violations of
local energy are permitted in quantum field theory. One powerful
approach is that of the quantum inequalities constraining the
magnitude  and duration of negative energy
regions\cite{L2,L3,BH1,FR,[10],EEF,[11a],[11],[12],[13],Yu,DV1,FT99,Fewster00,DV2,FV,EEF2,Pf,Yu2,FP,FM}.

The work on quantum inequalities was pioneered by Ford\cite{L2},
who derived an inequality type of bound on negative energy fluxes
for the quantized, massless, minimally-coupled scalar fields in
flat spacetime. Similar results for the sampled energy density
have been subsequently established for both massless and massive
scalar fields and electromagnetic fields in Minkowski
spacetime\cite{L3,[10]} as well as in static curved spacetimes
\cite{[11a],[11]}. However, in all these works, a Lorentzian
sampling function
\begin{equation}
f(\tau) = {\tau_0\over \pi (\tau^2 + \tau_0^2)}
\end{equation}
was employed in the calculations. Note that here $\tau_0$ sets the
characteristic averaging timescale.

Progress has been made toward removing the restriction of the
Lorentzian weight to include arbitrary sampling functions. In this
regard, Flanagan \cite{EEF} obtained optimal quantum inequalities
for the massless scalar field in two dimensions for arbitrary
smooth positive sampling functions. Fewster and colleagues
\cite{[13],FT99,Fewster00} derived the quantum inequalities for
the minimally-coupled scalar field in static curved spacetimes of
any dimension for an arbitrary sampling function. More recently,
Pfenning \cite{Pf} established a quantum inequality for
electromagnetic field in static curved spacetimes for arbitrary
positive  sampling functions using the techniques developed by
Fewster and colleagues for scalar fields in \cite{[13]} and
\cite{FT99}.

On the other hand, work is also being done for fields other than
scalar and electromagnetic ones.  Investigations on spin-${1\over
2}$ Dirac fields have been carried out by various authors
\cite{DV1,DV2,FV,Yu2,FM}. Specific quantum states with negative
energy density have been examined and shown to satisfy the quantum
inequalities for the scalar field obtained with a Lorentzian
sampling function\cite{DV1,Yu2}. Using arguments similar to those
of Flanagan's\cite{EEF}, Vollick \cite{DV2} derived an optimal
quantum inequality for the Dirac field in  two dimensions. Fewster
and Verch \cite{FV} have established the existence of quantum
inequalites for the Dirac (and Majorana) field in general
4-dimensional globally hyperbolic spacetimes, and more recently
Fewster and Mistry \cite{FM} have presented an explicit quantum
inequality bound for the Dirac field in four-dimensional Minkowski
spacetime using the modified methods for scalar fields. Recently
quantum inequalities have also been established for massive
spin-one Proca fields in globally hyperbolic spactimes whose
Cauchy surfaces are compact and have trivial first homology group
by Fewster and Pfenning \cite{FP}. As a further step along this
line, we will present a quantum inequality for massive
spin$-{3\over 2}$ Rarita-Schwinger fields in four-dimensional
Minkowski spacetime for arbitrary, smooth positive sampling
functions using the methods developed by Fewster and colleagues in
\cite{FV,FM}. The quantum inequality we are going to prove is, for
any real-valued, smooth, compactly supported function $g$,
\begin{equation}
\int dt\;\langle \rho (t, {\mathbf{x}})\rangle g(t)^2 \geq
-{1\over 12\pi^3}\;\int_m^{\infty}du\;|\widehat
g(u)|^2\;u^4Q_3^{RS}(u/m),
\end{equation}
where $\langle \rho (t, {\mathbf{x}})\rangle$ is the quantum
expectation value of the energy density of the field and
\begin{equation}
Q_3^{RS}(x) = 8 \biggl( 1 - {1\over x^2}
\biggr)^{3/2}-6Q^B_3(x)\;,
\end{equation}
and
\begin{equation}
Q^B_3(x) = \biggl(\; 1-{1\over x^2}\;\biggr)^{1/2} \biggl(\;
1-{1\over 2x^2}\;\biggr)-{1\over 2x^4} \ln (x +\sqrt{x^2-1})\;.
\end{equation}
We will work in the units where $c=\hbar=1$ and take the signature
of the metric to be $(+ ~- ~- ~-).$

\section{Rarita-Schwinger fields and the quantum inequality}

 Let us start with a review of the basics of free
Rarita-Schwinger fields\cite{RS,SK}. The Rarita-Schwinger fields
describe particles of spin ${3\over2}$ and they satisfy the
following equations
\begin{equation}
(-i\gamma\cdot\partial+m)\psi^{\mu}=0\;, \quad\quad
\gamma_{\mu}\psi^{\mu}=0\;,
\end{equation}
where $\gamma\cdot\partial=\gamma^{\mu}\partial_{\mu} $.  The
$\gamma$-matrices are given in terms of the Pauli matrices
$\sigma_{i}$ by
\begin{equation}
\gamma^{0}=\left(
\begin{array}{l l} 1 &\; 0  \\
0 & -1
\end{array}\right)\;, \qquad
\gamma^{i}=\left(
\begin{array}{l l}
\;\;0 & \sigma_{i} \\
-\sigma_{i} & 0
\end{array} \right)  \quad (i=1,2,3)
\end{equation}
and obey $\{ \gamma^{\mu}, \gamma^{\nu}  \} = 2 \eta^{\mu\nu}$.
The Lagrangian for the field can be written as
\begin{equation}
L={i\over
2}\bar{\psi^{\mu}}\gamma\cdot\tensor\partial\psi_{\mu}-m\bar{\psi^{\mu}}\psi_{\mu}+{i\over
6}\bar{\psi^{\mu}}(\gamma_{\mu}\tensor\partial_{\nu}+\gamma_{\nu}\tensor\partial_{\mu})\psi^{\nu}+{i\over
6}\bar{\psi^{\mu}}\gamma_{\mu}\gamma\cdot\tensor\partial\gamma^{\nu}\psi_{\nu}+{1\over
3}m\bar{\psi^{\mu}}\gamma_{\mu}\gamma^{\nu}\psi_{\nu}\;.
\end{equation}
A complete set of solutions for the field equations is given by
\begin{eqnarray}
\label{eq:Sol1}
 {\mathcal{U}}^{\mu}_{\bf{k}\sigma}e^{ik\cdot x}
\quad\quad \sigma=1,...4\;, \quad  \mu=0,1,2,3\;,
\end{eqnarray}
\begin{eqnarray}
\label{eq:Sol2} {\mathcal{V}}^{\mu}_{\bf{k}\sigma}e^{ik\cdot x}
\quad\quad \sigma=1,...4 \;, \quad \mu=0,1,2,3\;.
\end{eqnarray}
Here $ {\mathcal{U}}^{\mu}_{\bf{k}\sigma}$  and $
{\mathcal{V}}^{\mu}_{\bf{k}\sigma}$ can be expressed in terms of
the Dirac spinors  and a triad of four-vectors
$\epsilon_1({\bf{k}}), \epsilon_2({\bf{k}}), \epsilon_3({\bf{k}})
$ as
\begin{eqnarray}
{\mathcal{U}}_{{\bf{k}}1}&=&\epsilon_1({\bf{k}}) \otimes
u_{{\bf{k}}1}\;,
\\
{\mathcal{U}}_{{\bf{k}}2}&=&\sqrt{1\over 3}\;\epsilon_1({\bf{k}})
\otimes u_{{{\bf{k}}2}}
-\sqrt{2\over 3}\;\epsilon_3({\bf{k}}) \otimes u_{{\bf{k}}1}\;,\\
{\mathcal{U}}_{{\bf{k}}3}&=&\sqrt{1\over 3}\;\epsilon_2({\bf{k}})
\otimes
u_{{{\bf{k}}1}} +\sqrt{2\over 3}\;\epsilon_3({\bf{k}}) \otimes u_{{\bf{k}}2}\;,\\
{\mathcal{U}}_{{\bf{k}}4}&=&\epsilon_2({\bf{k}}) \otimes
u_{{\bf{k}}2}\;,
\end{eqnarray}
and
\begin{eqnarray}
{\mathcal{V}}_{{\bf{k}}1}&=&\epsilon_1({\bf{k}}) \otimes v_{{\bf{k}}1}\;,\\
{\mathcal{V}}_{{\bf{k}}2}&=&\sqrt{1\over 3}\;\epsilon_1({\bf{k}})
\otimes
v_{{{\bf{k}}2}} -\sqrt{2\over 3}\;\epsilon_3({\bf{k}})  \otimes v_{{\bf{k}}1}\;, \\
 {\mathcal{V}}_{{\bf{k}}3}&=&\sqrt{1\over 3}\;\epsilon_2({\bf{k}}) \otimes v_{{{\bf{k}}1}} +\sqrt{2\over 3}\;\epsilon_3({\bf{k}})
\otimes v_{{\bf{k}}2}\;,\\
{\mathcal{V}}_{{\bf{k}}4}&=&\epsilon_2({\bf{k}}) \otimes
v_{{\bf{k}}2}\;.
\end{eqnarray}
The  triad of four-vectors can be written as
\begin{equation}
\epsilon^{\mu}_i({\bf{k}}) =
L^{\mu}_{\nu}({\bf{k}})\epsilon^{\nu}_i(0)\;,
\end{equation}
where  $\epsilon^{\nu}_i(0)$  are given by
\begin{equation}
\epsilon_1(0)={1\over \sqrt{2}}\left(\begin{array}{l l l l}
0\\1\\i\\0\end{array}\right)\;, \quad
\epsilon_2(0)={1\over\sqrt{2}}\left(\begin{array}{l l l l}
\;0\\\;1\\-i\\\;0\end{array}\right)\;, \quad
\epsilon_3(0)=\left(\begin{array}{l l l l} {0}
\\0\\0\\1\end{array}\right)\;,
\end{equation}
and $L^{\mu}_{\nu}({\bf{k}})$ by \cite{SW}
\begin{eqnarray}
L^i_j({\bf{k}}) &=& \delta_{ij} + (\gamma-1)\hat k_i\hat
k_j\;,\nonumber\\
L^i_0({\bf{k}}) &=& L^0_i({\bf{k}}) = \hat k_i
\sqrt{\gamma^2-1}\;,\nonumber\\
L^0_0({\bf{k}}) &=& \gamma\;
\end{eqnarray}
with
\begin{equation}
\hat k_i \equiv { k_i\over |{\bf{k}}|}, \quad\quad \gamma \equiv
{\sqrt {{\bf{k}}^2 +m} \over m} = {\omega_{\bf{k}} \over m}\;.
\end{equation}
Let us note that if the momentum is taken to be along the $z$-axis
one has
\begin{equation}
\epsilon_1({\bf{k}})={1\over \sqrt{2}}\left(\begin{array}{l l l l}
0\\1\\i\\0\end{array}\right)\;, \quad
\epsilon_2({\bf{k}})={1\over\sqrt{2}}\left(\begin{array}{l l l l}
\;0\\\;1\\-i\\\;0\end{array}\right)\;, \quad
\epsilon_3({\bf{k}})=\left(\begin{array}{l l l l} {|{\bf{k}}|\over
m}
\\\;0\\\;0\\{\omega_{\bf{k}}\over m}\end{array}\right)\;.
\end{equation}
The Dirac spinors, $u_{{\bf{k}}\alpha}$ and  $v_{{\bf{k}}\alpha}$,
are
\begin{eqnarray}\label{eq:spinor1}
u_{{\bf{k}}\alpha}=\left(\begin{array}{l
l}{\sqrt{\frac{\omega_{\bf{k}}+m}{2\omega_{\bf{k}} V}}
\phi^\alpha}\\{\frac{\bf{\sigma}\cdot\bf k}
{\sqrt{2\omega_{\bf{k}}(\omega_{\bf{k}}+m)V}}\phi^\alpha}\end{array}\right),
\end{eqnarray}
\begin{eqnarray}\label{eq:spinor2}
v_{{\bf{k}}\alpha}=\left(\begin{array}{l
l}{\frac{\bf{\sigma}\cdot\bf
k}{\sqrt{2\omega_{\bf{k}}(\omega_{\bf{k}}+m)V}}\phi^\alpha}\\\sqrt{\frac{\omega_{\bf{k}}+m}{2\omega_{\bf{k}}
V}}\phi^\alpha\end{array}\right)
\end{eqnarray}
with $\phi^{1\dag}=(1,0)$, $ \phi^{2\dag}=(0,1)$.  Making use of
the above results, one can show that
\begin{equation}
\sum_{\sigma}{\mathcal{U}}^{\dagger\mu}_{\bf{k}\sigma}\;
{\mathcal{U}}_{\mu\bf{k}\sigma}=\sum_{\sigma}{\mathcal{V}}^{\dagger\mu}_{\bf{k}\sigma}\;
{\mathcal{V}}_{\mu\bf{k}\sigma}={4\over V}\;.
\end{equation}

 For spin ${3\over 2}$ fields, the canonical quantization
procedure becomes rather awkward, because of the difficulty of
separating dynamical degrees of freedom. In particular,
$\psi^{\mu}$ would have to be decomposed into its irreducible spin
${1\over 2}$ and spin ${3\over 2}$ parts and only the latter part
is subject to canonical quantization. To avoid the difficult
calculations which this entails, one can bypass the canonical
procedure altogether and work directly with the creation and
annihilation operators for the normal modes. A consistent
quantization can be obtained\cite{Lurie} by expanding the field in
terms of the complete set of solutions of Eq.~(\ref{eq:Sol1}) and
Eq.~(\ref{eq:Sol2})
\begin{eqnarray}
\psi^{\mu}(x)=\sum_{k}\sum^4_{\sigma=1}\Big[c_{\bf{k}\sigma}
{\mathcal{U}}^{\mu}_{\bf{k}\sigma}e^{ik\cdot x}+
d_{\bf{k}\sigma}^\dagger
{\mathcal{V}}^{\mu}_{\bf{k}\sigma}e^{-ik\cdot x}\Big],
\end{eqnarray}
and imposing  the following anticommutation relations on the
creation and annihilation operators:
\begin{equation}
\label{eq:Com1}
\{c_{\bf{k}\sigma}\;,\;c_{\bf{k'}\sigma'}^\dagger\}=\{d_{\bf{k}\sigma}\;,\;
d_{\bf{k'}\sigma'}^\dagger\}=\delta_{{\bf{k}}{\bf{k'}}}\delta_{\sigma\sigma'}\;,
\end{equation}

\begin{equation}
\label{eq:Com2}
\{c_{\bf{k}\sigma}\;,\;c_{\bf{k'}\sigma'}\}=\{d_{\bf{k}\sigma}\;,\;
d_{\bf{k'}\sigma'}\}=\{c_{\bf{k}\sigma}\;,\;d_{\bf{k'}\sigma'}^\dagger\}=\{c_{\bf{k}\sigma}\;,\;d_{\bf{k'}\sigma'}\}=0\;.
\end{equation}

In order to establish the quantum inequality for the
Rarita-Schwinger fields, we will use  the following symmetrized
energy momentum tensor $T^{\mu\nu}$ known as Belinfante tensor
\cite{Belinfante}.
\begin{eqnarray}
T^{\mu\nu}=&\partial^{\nu}&\bar\psi^{\alpha}{\partial L\over
\partial(\partial_{\mu}\bar\psi_{\alpha})}+{\partial L\over
\partial(\partial_{\mu}\psi_{\alpha})}\;\partial^{\nu}\psi^{\alpha}-g^{\mu\nu}L+{1\over2}\;\partial_{\beta}\biggl[\;{\partial L\over
\partial(\partial_{\beta}\psi_{\alpha})}\;J^{\mu\nu}\psi^{\alpha}+\bar\psi^{\alpha}\bar J^{\mu\nu}{\partial L\over
\partial(\partial_{\beta}\bar\psi_{\alpha})}\nonumber\\
&-&{\partial L\over
\partial(\partial_{\mu}\psi_{\alpha})}\;J^{\beta\nu}\psi^{\alpha}-\bar\psi^{\alpha}\bar J^{\beta\nu}{\partial L\over
\partial(\partial_{\mu}\bar\psi_{\alpha})}-{\partial L\over
\partial(\partial_{\nu}\psi_{\alpha})}\;J^{\beta\mu}\psi^{\alpha}-\bar\psi^{\alpha}\bar J^{\beta\mu}{\partial L\over
\partial(\partial_{\nu}\bar\psi_{\alpha})}\;
\biggr]\;,
\end{eqnarray}
where $J^{\mu\nu}$ and $\bar J^{\mu\nu}$ are the generators of the
Lorentz transformations for $\psi$ and $\bar\psi$ respectively.
Consider an infinitesimal Lorentz transformation
\begin{equation}
x^{'\mu}=x^{\mu}+\omega^{\mu\nu}x_{\nu}\;.
\end{equation}
We have
\begin{eqnarray}
\delta \psi &=&{1\over 2}\omega_{\alpha\beta} \bigl(
S^{\alpha\beta}+L^{\alpha\beta}\bigr)\psi\;,\\
\delta \bar\psi &=&{1\over 2}\omega_{\alpha\beta} \bigl(
S^{\alpha\beta}-L^{\alpha\beta}\bigr)\psi\;,
\end{eqnarray}
where $S^{\alpha\beta}$ and $L^{\alpha\beta}$, which operate on
the spacetime vector and internal spinor indices of $\psi$
respectively, are given by
\begin{eqnarray}
(S^{\alpha\beta})^{\mu\nu}&=&\eta^{\alpha\mu}\eta^{\beta\nu}-\eta^{\alpha\nu}\eta^{\beta\mu}\;,\\
L^{\alpha\beta}&=&{1\over4}\;\bigl[\;\gamma^{\alpha},\gamma^{\beta}\;\bigr]\;.
\end{eqnarray}
So, it follows that
\begin{equation}
J^{\alpha\beta}=S^{\alpha\beta}+L^{\alpha\beta}\;,\quad\quad \bar
J^{\alpha\beta}=S^{\alpha\beta}-L^{\alpha\beta}\;.
\end{equation}
The energy momentum tensor is obtained,  after a concrete
calculation using the above results and taking the equations of
motion into account as
\begin{equation}
T_{\mu\nu}={i\over 4}\;\bigl[\;
\bar\psi^{\rho}\gamma_{\mu}\tensor\partial_{\nu}\psi_{\rho} +
\bar\psi^{\rho}\gamma_{\nu}\tensor\partial_{\mu}\psi_{\rho}
\;\bigr]\;.
\end{equation}
Hence the energy density is
\begin{equation}
T_{00}={i\over 2}\;\bigl[\; \psi^{\dagger\rho}\dot\psi_{\rho}-
\dot\psi^{\dagger\rho}\psi_{\rho} \;\bigr]\;.
\end{equation}
The renormalized expectation value of the energy density, i.e.,
$\langle\rho\rangle=\langle:T_{00}:\rangle$, in an arbitrary
quantum state, is given by
\begin{eqnarray}
\langle \rho (t, {\mathbf{x}})\rangle={1\over
2}\sum_{\mathbf{k,k'}}\sum_{\sigma\sigma'}&&\biggr\{
(\omega_{\mathbf{k}} + \omega_{\mathbf{k'}} )\big[\; \langle
c^\dagger_{\bf{k}\sigma}\; c_{\bf{k'}\sigma'}\rangle
{\mathcal{U}}^{\dagger\mu}_{\bf{k}\sigma}\;
{\mathcal{U}}_{\mu\bf{k'}\sigma'} e^{i(k-k')\cdot x}\nonumber\\
&&+ \langle d^\dagger_{\bf{k}\sigma}\; d_{\bf{k'}\sigma'}\rangle
{\mathcal{V}}^{\dagger\mu}_{\bf{k}\sigma}\;
{\mathcal{V}}_{\mu\bf{k'}\sigma'} e^{-i(k-k')\cdot
x}\;\big]\nonumber\\
&&+(\omega_{\mathbf{k'}} - \omega_{\mathbf{k}} )\bigl[\; \langle
d_{\bf{k}\sigma}\; c_{\bf{k'}\sigma'}\rangle
{\mathcal{V}}^{\dagger\mu}_{\bf{k}\sigma}\;
{\mathcal{U}}_{\mu\bf{k'}\sigma'} e^{-i(k+k')\cdot x}\nonumber\\
&&- \langle c^\dagger_{\bf{k}\sigma}\;
d^\dagger_{\bf{k'}\sigma'}\rangle
{\mathcal{U}}^{\dagger\mu}_{\bf{k}\sigma}\;
{\mathcal{V}}_{\mu\bf{k'}\sigma'} e^{i(k+k')\cdot x}\; \bigr]
 \biggr\}\;.
\end{eqnarray}
Consider the sampled energy density measured by a stationary
observer at the spatial origin
\begin{equation}
\langle \rho \rangle_f=\int^{\infty}_{-\infty}\;\langle \rho (t,
{\mathbf{0}})\rangle f(t)\;dt\;,
\end{equation}
where $f$ is a non-negative sampling function. Then
\begin{eqnarray}
\langle \rho \rangle_f={1\over
2}\sum_{\mathbf{k,k'}}\sum_{\sigma\sigma'}&&\biggr\{
(\omega_{\mathbf{k}} + \omega_{\mathbf{k'}} )\bigl[\; \langle
c^\dagger_{\bf{k}\sigma}\; c_{\bf{k'}\sigma'}\rangle
{\mathcal{U}}^{\dagger\mu}_{\bf{k}\sigma}\;
{\mathcal{U}}_{\mu\bf{k'}\sigma'} \widehat{f}(\omega_{\mathbf{k'}}
- \omega_{\mathbf{k}})\nonumber\\
&&+\langle d^\dagger_{\bf{k}\sigma}\; d_{\bf{k'}\sigma'}\rangle
{\mathcal{V}}^{\dagger\mu}_{\bf{k}\sigma}\;
{\mathcal{V}}_{\mu\bf{k'}\sigma'} \widehat{f}(\omega_{\mathbf{k}}
-
\omega_{\mathbf{k'}})\;\bigr]\nonumber\\
&&+(\omega_{\mathbf{k'}} - \omega_{\mathbf{k}} )\bigl[\; \langle
d_{\bf{k}\sigma}\; c_{\bf{k'}\sigma'}\rangle
{\mathcal{V}}^{\dagger\mu}_{\bf{k}\sigma}\;
{\mathcal{U}}_{\mu\bf{k'}\sigma'} \widehat{f}(\omega_{\mathbf{k}}
+ \omega_{\mathbf{k'}})\nonumber\\
&&-\langle c^\dagger_{\bf{k}\sigma}\;
d^\dagger_{\bf{k'}\sigma'}\rangle
{\mathcal{U}}^{\dagger\mu}_{\bf{k}\sigma}\;
{\mathcal{V}}_{\mu\bf{k'}\sigma'} \widehat{f}(-\omega_{\mathbf{k}}
- \omega_{\mathbf{k'}})\; \bigr]
 \biggr\}\;,
\end{eqnarray}
where $\widehat{f}$ is the Fourier transform of $f$ defined by
\begin{equation}
\widehat{f}(\omega)=\int^{\infty}_{-\infty}\;f(t) e^{-i\omega
t}dt\;.
\end{equation}
Let $f=g^2$ and define a family of operators by
\begin{equation}
{\mathcal{O}}^{\mu}_{\lambda
i}=\sum_{{\mathbf{k'}},\sigma'}\biggl\{\overline{\widehat{g}(-\omega_{\mathbf{k'}}+\lambda)}\;c_{\bf{k'}\sigma'}
{\mathcal{U}}^{\mu}_{{\bf{k'}}\sigma' i} +
\overline{\widehat{g}(\omega_{\mathbf{k'}}+\lambda)}\;d^\dagger_{\bf{k'}\sigma'}
{\mathcal{V}}^{\mu}_{{\bf{k'}}\sigma' i} \biggr\}\;,
\end{equation}
where $i=1,...4$ is the spinor index and $\overline{\widehat{g}}$
is the conjugate of the Fourier transform. Using the
anticommutation relations and the fact that
\begin{equation}
\sum_{\sigma}{\mathcal{U}}^{\dagger\mu}_{\bf{k}\sigma}\;
{\mathcal{U}}_{\mu\bf{k}\sigma}=\sum_{\sigma}{\mathcal{V}}^{\dagger\mu}_{\bf{k}\sigma}\;
{\mathcal{V}}_{\mu\bf{k}\sigma}={4\over V}\;,
\end{equation}
we find
\begin{eqnarray}
{\mathcal{O}}^{\dagger\mu}_{\lambda i}{\mathcal{O}}_{\mu_\lambda
i}=S_{\lambda}+\sum_{{\mathbf{k}}{\mathbf{k'}}}\sum_{\sigma\sigma'}
&&
 \biggl\{\widehat{g}(-\omega_{\mathbf{k}}+\lambda)
\overline{\widehat{g}(-\omega_{\mathbf{k'}}+\lambda)}\;
c^\dagger_{\bf{k}\sigma} c_{\bf{k'}\sigma'}
{\mathcal{U}}^{\dagger\mu}_{{\bf{k}}\sigma }
{\mathcal{U}}_{\mu\bf{k'}\sigma' }
\nonumber\\
&&
-\widehat{g}(\omega_{\mathbf{k}}+\lambda)\overline{\widehat{g}(\omega_{\mathbf{k'}}+\lambda)}\;d^\dagger_{\bf{k'}\sigma'}
d_{\bf{k}\sigma}
{\mathcal{V}}^{\dagger\mu}_{{\bf{k}}\sigma}{\mathcal{V}}_{\mu\bf{k'}\sigma'}\nonumber\\
&& +\widehat{g}(\omega_{\mathbf{k}}+\lambda)
\overline{\widehat{g}(-\omega_{\mathbf{k'}}+\lambda)}\;
d_{\bf{k}\sigma} c_{\bf{k'}\sigma'}
{\mathcal{V}}^{\dagger\mu}_{{\bf{k}}\sigma }
{\mathcal{U}}_{\mu\bf{k'}\sigma' }
\nonumber\\
&& +\widehat{g}(-\omega_{\mathbf{k}}+\lambda)
\overline{\widehat{g}(\omega_{\mathbf{k'}}+\lambda)}\;
c^\dagger_{\bf{k}\sigma} d^\dagger_{\bf{k'}\sigma'}
{\mathcal{U}}^{\dagger\mu}_{{\bf{k}}\sigma }
{\mathcal{V}}_{\mu\bf{k'}\sigma' }
 \biggr\}\;,
\end{eqnarray}
where
\begin{equation}
S_{\lambda}= {4\over V}
\sum_{{\mathbf{k}}}\;|\widehat{g}(\omega_{\mathbf{k}}+\lambda)|^2\;.
\end{equation}
 Making use of the following relation which was proven by
Fewster and colleagues \cite{FV,FM} for real-valued, smooth,
compactly supported $g=f^{1/2}$
\begin{equation}
(\omega+\omega')\widehat{f}(\omega-\omega') =
\int^\infty_{-\infty}\; {\lambda\over
\pi}\widehat{g}(\omega-\lambda)\overline{\widehat{g}(\omega'-\lambda)}\;d\lambda\;,
\end{equation}
we can show that
\begin{equation}
\langle \rho \rangle_f = {1\over 2\pi}\;\int^\infty_{-\infty}\;
(\langle {\mathcal{O}}^{\dagger\mu}_{\lambda
i}{\mathcal{O}}_{\mu_\lambda i}\rangle-S_{\lambda} ) \lambda \;
d\lambda\;.
\end{equation}
Now let us calculate the anticommutator $\{
{\mathcal{O}}^{\dagger\mu}_{\lambda i}\;,
{\mathcal{O}}_{\mu_\lambda i}\}$ to get
\begin{eqnarray}
\{{\mathcal{O}}^{\dagger\mu}_{\lambda
i}\;,{\mathcal{O}}_{\mu_\lambda i} \}
&=&\sum_{{\mathbf{k}}{\mathbf{k'}}}\sum_{\sigma\sigma'}
 \biggl\{\widehat{g}(-\omega_{\mathbf{k}}+\lambda)
\overline{\widehat{g}(-\omega_{\mathbf{k'}}+\lambda)}\;
\{c^\dagger_{\bf{k}\sigma}\;, c_{\bf{k'}\sigma'}\}
{\mathcal{U}}^{\dagger\mu}_{{\bf{k}}\sigma }
{\mathcal{U}}_{\mu\bf{k'}\sigma' }
\nonumber\\
&&\quad\quad\quad
-\widehat{g}(\omega_{\mathbf{k}}+\lambda)\overline{\widehat{g}(\omega_{\mathbf{k'}}+\lambda)}\;\{d^\dagger_{\bf{k'}\sigma'}
\;, d_{\bf{k}\sigma}\}
{\mathcal{V}}^{\dagger\mu}_{{\bf{k}}\sigma}{\mathcal{V}}_{\mu\bf{k'}\sigma'}\biggr\}\nonumber\\
&=&\sum_{{\mathbf{k}}}\;|\widehat{g}(-\omega_{\mathbf{k}}+\lambda)|^2\;\sum_{\sigma}{\mathcal{U}}^{\dagger\mu}_{\bf{k}\sigma}\;
{\mathcal{U}}_{\mu\bf{k}\sigma} +
\sum_{{\mathbf{k}}}\;|\widehat{g}(\omega_{\mathbf{k}}+\lambda)|^2\;\sum_{\sigma}{\mathcal{V}}^{\dagger\mu}_{\bf{k}\sigma}\;
{\mathcal{V}}_{\mu\bf{k}\sigma}\nonumber\\
&=& {4\over V}
\sum_{{\mathbf{k}}}\;|\widehat{g}(-\omega_{\mathbf{k}}-(-\lambda)
)|^2+{4\over V}
\sum_{{\mathbf{k}}}\;|\widehat{g}(\omega_{\mathbf{k}}+\lambda)|^2\nonumber\\
&=& {4\over V}
\sum_{{\mathbf{k}}}\;|\widehat{g}(\omega_{\mathbf{k}}-\lambda
)|^2+{4\over V}
\sum_{{\mathbf{k}}}\;|\widehat{g}(\omega_{\mathbf{k}}+\lambda)|^2\nonumber\\
&=& S_{-\lambda}+S_{\lambda}\;.
\end{eqnarray}
Here we have used the anticommutation relations
Eqs.~(\ref{eq:Com1}, \ref{eq:Com2}) and appealed to the fact that
$|\widehat{g}(x)|$ is an even function since $g$ is real. An
application of the above result leads to
\begin{eqnarray}
\langle \rho \rangle_f &=& {1\over 2\pi}\;\int^\infty_{0}\;
(\langle{\mathcal{O}}^{\dagger\mu}_{\lambda
i}{\mathcal{O}}_{\mu_\lambda i}\rangle-S_{\lambda} ) \lambda \;
d\lambda + {1\over 2\pi}\;\int^0_{-\infty}\;
\biggl[(S_{\lambda}+S_{-\lambda})-\langle{\mathcal{O}}^{\mu}_{\lambda
i}{\mathcal{O}}^{\dagger}_{\mu_\lambda i}\rangle-S_{\lambda}
\biggr]
\lambda \; d\lambda \nonumber\\
&=& {1\over 2\pi}\;\int^\infty_{0}\;
(\langle{\mathcal{O}}^{\dagger\mu}_{\lambda
i}{\mathcal{O}}_{\mu_\lambda i}\rangle-S_{\lambda} ) \lambda \;
d\lambda + {1\over 2\pi}\;\int^0_{-\infty}\;
(S_{-\lambda}-\langle{\mathcal{O}}^{\mu}_{\lambda
i}{\mathcal{O}}^{\dagger}_{\mu_\lambda i}\rangle) \lambda \;
d\lambda\;.
\end{eqnarray}
For all quantum states in which $
\langle{\mathcal{O}}^{\dagger\mu}_{\lambda
i}{\mathcal{O}}_{\mu_\lambda i}\rangle \geq 0$, the following
inequality holds
\begin{eqnarray}
\langle \rho \rangle_f &\geq & -{1\over 2\pi}\;\int^\infty_{0}\;
 \lambda S_{\lambda}  \; d\lambda + {1\over 2\pi}\;\int^0_{-\infty}\;
 \lambda S_{-\lambda}  \;
d\lambda\nonumber\\
& =& -{1\over \pi}\;\int^\infty_{0}\; \lambda  S_{\lambda}  \;
d\lambda  = -{4\over \pi}\;\int^\infty_{0}\;d\lambda \; \lambda
{1\over V} \sum_{{\mathbf{k}}}\;|\widehat{g}(\omega_{\mathbf{k}} +
\lambda )|^2 \; .
\end{eqnarray}
Taking the continuum limit $ {1\over V}
\sum_{\mathbf{k}}\rightarrow \int  { d^3 {\mathbf{k}}\over
(2\pi)^3}$ and following the same steps as in Ref.~\cite{FM}, we
can show that
\begin{equation}
\langle \rho \rangle_f \geq  - {1\over \pi^3}\;\int_m^{\infty}\;
du \;|\widehat g(u)|^2\;\left(\;{2\over 3}u(u^2-m^2)^{3/2}-{1\over
2}u^4Q^B_3\biggl({u\over m}\biggr) \right)\;,
\end{equation}
where
\begin{equation}
Q^B_3(x) = \biggl(\; 1-{1\over x^2}\;\biggr)^{1/2} \biggl(\;
1-{1\over 2x^2}\;\biggr)-{1\over 2x^4} \ln (x +\sqrt{x^2-1})\;.
\end{equation}
Using the translational invariance of the theory, the quantum
inequality can be expressed by the sampled energy density measured
by a stationary observer at an arbitrary space time point as
\begin{equation}
\int\; dt\;\langle \rho (t, {\mathbf{x}})\rangle g(t)^2 \geq
-{1\over 12\pi^3}\;\int_m^{\infty}\;du\;|\widehat
g(u)|^2\;u^4Q_3^{RS}(u/m).
\end{equation}
Here
\begin{equation}
Q_3^{RS}(x) = 8 \biggl( 1 - {1\over x^2}
\biggr)^{3/2}-6Q^B_3(x)\;.
\end{equation}
The bound is finite since $|\widehat g(u)|^2$ decays faster than
any polynomial in $u$ and $u^4Q_3^{RS}(u/m)$ grows like $u^4$ as
$u\rightarrow \infty$. Comparing the above result with that
obtained by  Fewster and Mistry \cite{FM} for the Dirac field,
i.e.,
\begin{equation}
\int\; dt\;\langle \rho (t, {\mathbf{x}})\rangle g(t)^2 \geq
-{1\over 12\pi^3}\;\int_m^{\infty}\;du\;|\widehat
g(u)|^2\;u^4Q_3^{D}(u/m)\;,
\end{equation}
one can see that
\begin{equation}
Q_3^{RS}(x) = 2 Q_3^{D}(x)\;.
\end{equation}
So the quantum inequality bound for the free massive
Rarita-Schwinger field is weaker, by a factor of 2, than that of
the Dirac field.

\section{discussions}

In conclusion, we have derived a quantum inequality for the free
massive  Rarita-Schwinger field in Minkowski spacetime for
arbitrary smooth positive sampling functions following methods
developed by Fewster and colleagues \cite{FV,FM}. Our quantum
inequality bound for Rarita-Schwinger fields is weaker, by a
factor of 2, than that for the Dirac field. This seems to be a
result of the fact that massive Rarita-Schwinger fields have twice
as many number of field degrees of freedom  as the Dirac fields.
Recall the quantum inequalities that have been established for
quantized fields of integer spin in four dimensional Minkowski
spacetime using similar methods \cite{[13],FT99,Pf,FP}
\begin{equation}
\int\; dt\;\langle \rho (t, {\mathbf{x}})\rangle g(t)^2 \geq
-{{\mathcal{S}}\over 16\pi^3}\;\int_m^{\infty}\;du\;|\widehat
g(u)|^2\;u^4Q^B_3(u/m)\;,
\end{equation}
where ${\mathcal{S}}$ is just the number of the field degrees of
freedom. ${\mathcal{S}} = 1$ for scalar fields (spin zero), 2 for
electromagnetic fields (spin 1) and 3 for massive Proca fields
(spin 1). Note, however,  that in general curved spacetimes the
quantum inequalities of these theories may not simply related by
an overall factor. In the same spirit, quantum inequalities
obtained so far for the half-integral spin fields can also be cast
into the following unified form
\begin{equation}
\int\; dt\;\langle \rho (t, {\mathbf{x}})\rangle g(t)^2 \geq
-{{\mathcal{S}}\over 24\pi^3}\;\int_m^{\infty}\;du\;|\widehat
g(u)|^2\;u^4Q^{F}_3(u/m)\;,
\end{equation}
where
\begin{equation}
Q^{F}_3(x)= 4 \biggl( 1 - {1\over x^2} \biggr)^{3/2}-3Q^B_3(x)\;.
\end{equation}
Here ${\mathcal{S}} = 2$ for Dirac fields and 4 for
Rarita-Schwinger fields.

An interesting point to note from the above results is that,
separately for fields of integer spin (bosonic fields) and those
of half-integral spin (fermionic fields), the quantum inequality
bound gets weaker as the the number of degrees of freedom of the
field increases. However, since none of these bounds are optimal,
this observation is now more a conjecture than a conclusion.
Optimal bounds for all these fields have to be found to see if
this is true or even if it is true for both bosonic and fermionic
fields combined. Nevertheless, we would like to point out that
this is physically plausible, since the more the number of field
degrees of freedom, the more freedom one has to create negative
energy, therefore, the weaker the quantum inequality bound ought
be.

\begin{acknowledgments}
We would like to acknowledge the support  by the National Natural
Science Foundation of China  under Grants No. 10075019 and No.
10375023.
\end{acknowledgments}


\end{document}